\documentclass[journal]{IEEEtran}
\usepackage[noadjust]{cite}
\usepackage{srcltx}
\usepackage{graphicx}
\usepackage{epstopdf}
\usepackage{psfrag}
\usepackage{epsfig}
\usepackage[tbtags,sumlimits,nointlimits,reqno]{amsmath}
\usepackage{amssymb}
\usepackage{xcolor}
\usepackage{float}
\usepackage{subfigure}
\usepackage{soul}
\usepackage{footnote}
\usepackage{amsmath}
\usepackage{color}

\usepackage[framemethod=TikZ]{mdframed}
\usepackage{lipsum}
\mdfdefinestyle{MyFrame}{%
    linecolor=blue,
    outerlinewidth=2pt,
    roundcorner=20pt,
    innertopmargin=\baselineskip,
    innerbottommargin=\baselineskip,
    innerrightmargin=20pt,
    innerleftmargin=20pt,
    backgroundcolor=gray!50!white}

\begin{document}
\title{Mixer-Duplexer-Antenna Leaky-Wave System \\ Based on Periodic Space-Time Modulation}
\author{Sajjad Taravati and Christophe Caloz,~\IEEEmembership{Fellow,~IEEE}
\thanks{S. Taravati and C. Caloz are with the Department
of Electrical Engineering and Poly-Grames research center, \'{E}cole Polytechnique de Montr\'{e}al, Montr\'{e}al, Qu\'{e}bec, H3T 1J4, Canada (e-mail: sajjad.taravati@polymtl.ca).}%
\thanks{Manuscript received ***, 2016; revised ***, 2016.}}

\markboth{Transaction on Antennas and Propagation,~Vol.~*, No.~*, ***~2016}%
{*** \MakeLowercase{\textit{et al.}}: Bare Demo of IEEEtran.cls for IEEE Journals}

\maketitle

\begin{abstract}
We present a mixer-duplexer-antenna leaky-wave
system based on periodic space-time modulation. This system operates as a full transceiver, where the upconversion and downconversion mixing operations are accomplished via space-time transitions, the duplexing operation is induced by the nonreciprocal nature of the structure, and the radiation operation is provided by the leaky-wave nature of the wave. A rigorous electromagnetic solution is derived for the dispersion relation and field distributions. The system is implemented in the form of a spatio-temporally modulated microstrip leaky-wave structure incorporating an array of sub-wavelengthly spaced varactors modulated
by a harmonic wave. In addition to the overall mixer-duplexer-antenna operation, frequency beam scanning at fixed input frequency is demonstrated as one of the interesting features of the system. A prototype is realized and demonstrated by full-wave and experimental results.
\end{abstract}

\begin{IEEEkeywords}
Leaky-wave antenna, space-time modulation, nonreciprocity, duplexer, mixer, transceiver.
\end{IEEEkeywords}

\IEEEpeerreviewmaketitle

\section{Introduction}

Space-time modulated structures, where the electric or magnetic properties of the medium are periodically modulated in space and time, have been theoretically studied for a long time as travelling-wave parametric amplifiers~\cite{Cullen_NAT_1958,Tien_JAP_1958,Oliner_1959,Oliner_PIEEE_1963,Cassedy_PIEEE_1965,Cassedy_PIEEE_1967,Peng1969,Chu1969,Chu1972}.
Recently, it has been pointed out that such unidirectionally modulated structures are fundamentally nonreciprocal~\cite{Fan_NPH_2009}, and this has triggered a regain of interest in space-time modulated systems~\cite{Fan_NPH_2009,Kalluri_2010,Fan_PRL_108_2012,Fan_PRL_109_2012,Engheta_APL_2014,Wang_TMTT_2014}. Several applications of space-time modulated nonreciprocal structures have been subsequently reported, including isolators~\cite{Fan_NPH_2009}, circulators~\cite{Sounas_NATCOM_09_2013,Alu_NP_2014}, nonreciprocal metasurfaces~\cite{Alu_PRB_2015,Shalaev_OME_2015,Fan_APL_2016,Taravati_ArXiv_2016} and nonreciprocal antennas~\cite{Taravati_APS_2015,Hadad_URSI_2015,Alu_AWPL_2015,Alu_TAP_63_2015,Alu_PNAS_2016}.

This paper presents a two-port nonreciprocal space-time modulated leaky-wave system which simultaneously performs the tasks of mixing, duplexing and radiation, hence operating as a complete transceiver system. Most of previously reported nonreciprocal leaky-wave systems were based on magnetically biased ferrites~\cite{Buchta_1964,Kodera_TMTT_04_2009,Tsutsumi_TAP_2009,Kodera_AWPL_11_2009,Kodera_TAP_10_2010,Kodera_AWPL_01_2012,Kodera_AWPL_01_2012,Volakis_TAP_2013,Volakis_TAP_2014}, and hence suffer from the drawbacks inherent to ferrite technology, namely bulkiness, heaviness, incompatibility with integrated circuits and high cost, in addition to requiring excessive bias field beyond the X-band for resonance-based components~\cite{Lax_1962}. The first nonreciprocal leaky-wave system based on space-time modulation, and hence requiring no biasing magnet, was independently proposed at the same time in~\cite{Taravati_APS_2015} and~\cite{Hadad_URSI_2015}, and the latter proposal was extended to an experimental demonstration in~\cite{Alu_PNAS_2016}.

As~\cite{Alu_PNAS_2016}, this paper, which is an extension of~\cite{Taravati_APS_2015}, presents a nonreciprocal space-time modulated leaky-wave system. However, this work and~\cite{Taravati_APS_2015} feature fundamental differences. The system in~\cite{Alu_PNAS_2016} is a single-port leaky-wave structure where the input frequency is up-converted to radiate whereas an incoming wave at the radiation frequency is absorbed in the structure and does not reach the input port. In contrast, this paper presents a two-port structure that performs the operation of a full transceiver system, with uplink and downlink space-time transitions, representing upconversion and downconversion mixing, and separation of the uplink and downlink paths, representing duplexing. Moreover, this paper demonstrates space-time frequency beam scanning at fixed input frequency by variation of a modulation parameter. Finally, this paper presents a detailed electromagnetic resolution of the problem for the dispersion relation and field structure.

The paper is organized as follows. Section~\ref{sec:S-T_mod_LWM} presents the operation principle and analytical solution of the nonreciprocal space-time modulated leaky-wave structure. Section~\ref{sec:Mix_Dupl_LWA} proposes a practical realization of the corresponding mixer-duplexer-antenna system, based on a half-wavelength microstrip leaky-wave antenna incorporating subwavelengthly-spaced modulating varactors. Section~\ref{sec:prototype} describes the implementation of the system and characterizes it in terms of its dispersion relation and field distributions. Full-wave and experimental results are provided in Sec.~\ref{sec:exp_res}. Finally, Section~\ref{sec:conc} concludes the paper.

\section{Space-time Modulated Leaky-Wave Structure}\label{sec:S-T_mod_LWM}

\subsection{Operation Principle}

\begin{figure}[ht]
\vspace{2cm}
\centering
\psfrag{m}[c][l][0.8]{\shortstack{transmitted wave:\\$\omega_\text{0}$}}
\psfrag{b}[c][c][0.8]{\shortstack{space-time modulated\\leaky-wave structure}}
\psfrag{B}[c][c][0.8][31]{$\epsilon(z,t)=\epsilon_\text{e}(1+\delta_\text{m} \cos(\omega_\text{m}t-\beta_\text{m}z))$}
\psfrag{d}[c][c][0.8]{\shortstack{radiated wave:\\$\omega_\text{1}=\omega_\text{0}+\omega_\text{m}$}}
\psfrag{D}[c][c][0.8]{\shortstack{incoming wave:\\$\omega_\text{1}$}}
\psfrag{l}[l][c][0.8]{$L$}
\psfrag{x}[l][c][0.8]{$x$}
\psfrag{z}[l][c][0.8]{$z$}
\psfrag{a}[l][c][0.8]{\shortstack{received wave:\\$\omega_\text{0}=\omega_\text{1}-\omega_\text{m}$}}
\centering{\includegraphics[width=0.9\columnwidth] {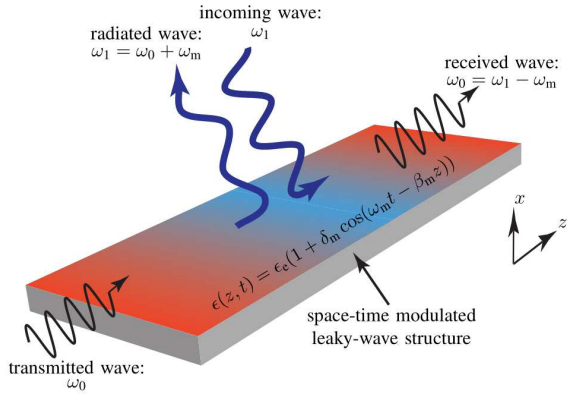}}
\vspace{-2.9cm}
\caption{Generic representation of a periodically space-time modulated leaky-wave structure, consisting of a medium with permittivity $\epsilon(z,t)=\epsilon_\text{e}(1+\delta_\text{m}\cos(\omega_\text{m}t-\beta_\text{m}z))$ interfaced with air. The wave transmitted at the left port is frequency up-converted and radiated under a specified angle (uplink), while a wave incoming under the same angle and at the same frequency is down-converted back to the original frequency and received at the right port (downlink). The system is inherently nonreciprocal due to the directionality of the space-time modulation.}
\label{Fig:ST_LW_med}
\end{figure}

Figure~\ref{Fig:ST_LW_med} shows the generic representation of a periodically space-time modulated leaky-wave structure. The structure consists of a medium with permittivity
\begin{equation}
\epsilon(z,t)= \epsilon_\text{e}(1+\delta_\text{m} \cos(\omega_\text{m}t-\beta_\text{m}z))
\label{eqa:S-T_perm}
\end{equation}
interfaced with air. In Eq.~\eqref{eqa:S-T_perm}, $\epsilon_\text{e}$ is the effective permittivity of the unmodulated medium, $\delta_\text{m}$ is the modulation depth, and $\omega_\text{m}$ and $\beta_\text{m}$ are the modulation temporal and spatial frequencies, respectively. Due to the directionally of the space-time modulation, which propagates in the $+z$ direction with velocity $v_\text{m}=\omega_\text{m}/\beta_\text{m}$, the structure is inherently nonreciprocal. The system has two ports, which support a transmitted wave and a received wave in an uplink/downlink transceiver scenario.

\begin{figure}
\centering
\psfrag{A}[c][l][0.8]{(a)}
\psfrag{B}[c][c][0.8]{(b)}
\psfrag{a}[c][l][0.8]{$\omega$}
\psfrag{b}[c][c][0.8]{$\beta$}
\psfrag{c}[c][c][0.8]{$\beta_\text{0}$}
\psfrag{d}[c][c][0.8]{$\beta_\text{1}$}
\psfrag{e}[c][c][0.8]{$\omega_\text{0}$}
\psfrag{f}[c][c][0.8]{$\omega_\text{1}$}
\psfrag{g}[c][c][0.8]{$\beta_\text{m}$}
\psfrag{h}[c][c][0.8]{$\omega_\text{m}$}
\psfrag{i}[c][l][0.8]{\shortstack{guided-mode\\(modulation)}}
\psfrag{j}[l][l][0.8]{\shortstack{air line}}
\psfrag{k}[c][c][0.8]{\shortstack{first higher-order\\-leaky-mode}}
\psfrag{l}[c][c][0.8]{\shortstack{no mode\\for transition}}
\psfrag{C}[c][c][0.8]{$-\beta_\text{0}$}
\psfrag{D}[c][c][0.8]{$-\beta_\text{1}$}
\psfrag{s}[c][c][0.8]{\shortstack{space-time\\transition}}
\psfrag{z}[c][c][0.8]{radiation region}
\centering{\includegraphics[width=0.8\columnwidth] {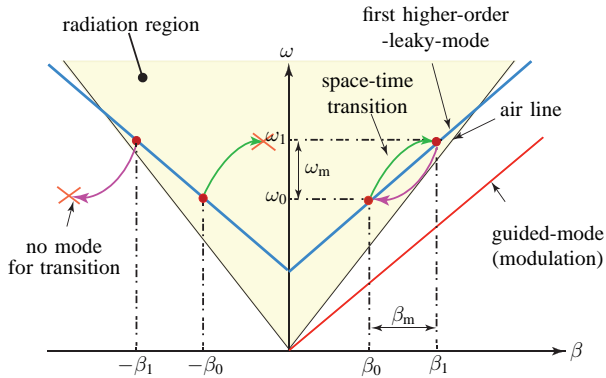}}
\caption{Schematic dispersion diagram explaining the nonreciprocal uplink and downlink space-time transitions in the leaky-wave structure of Fig.~\ref{Fig:ST_LW_med}.}
\label{Fig:disp_diag}
\end{figure}

Figure~\ref{Fig:disp_diag} qualitatively explains the operation principle of the space-time modulated leaky-wave structure in Fig.~\ref{Fig:ST_LW_med}. The space-time modulated permittivity is provided by injection of a harmonic wave in a guided-mode of the structure, as will be practically shown in Sec.~\ref{sec:Mix_Dupl_LWA}, while the wave of interest is supported by a leaky-mode of the structure, where mixing with the space-time modulation induces nonreciprocal uplink and downlink oblique transitions, represented by the green arrow and the blue arrow, respectively, in Fig.~\ref{Fig:disp_diag}. The exact operation is as follows.

In the uplink, a signal wave with frequency $\omega_\text{0}$ is injected into the transmitter port and propagates in the $+z$ direction, corresponding to the right-hand side of the dispersion diagram in Fig.~\ref{Fig:disp_diag}, as $E_\text{t}=E_0 e^{j (\omega_\text{0} t-\beta_\text{0} z)}$. As a result of mixing with the periodic modulation, this wave experiences progressive temporal frequency transition (up-conversion) from $\omega_\text{0}$ to $\omega_\text{1}=\omega_\text{0}+\omega_\text{m}$ along with spatial frequency transition from $\beta_\text{0}$ to $\beta_\text{1}=\beta_\text{0}+\beta_\text{m}$\footnote{The physical explanation for the generation of new frequencies due to medium temporal variations -- specifically vertical transitions in the dispersion diagram -- has been given in several texts, such as for instance~\cite{Kalluri_2010}. In the particular case of periodic temporal variations, one may draw an analogy with periodic spatial variations in a one-dimensional spatial periodic structure (or electromagnetic bandgap material). As a result of complex scattering from the spatial periodic modulation, the field solution to Maxwell equations in the structure is spatially non-sinusoidal (e.g. having more or less energy concentrated in the high or low refractive index regions for the lower and higher bands, respectively~\cite{Joannopoulos_2nd_ed}), except in the long-wavelength regime where homogeneization applies and leads to a well defined refractive index. Being non-sinusoidal, this waveform is actually a superposition of an infinite number of Floquet space harmonics, $\beta_0+n\beta_\text{m}$, corresponding to horizontal transitions in the dispersion diagram. Similarly, in the case of a temporal periodic modulation, waves get scattered forward and backward~\cite{Kalluri_2010} to form a temporally non-sinusoidal waveform, corresponding to an infinite number of Floquet \emph{time} harmonics, i.e. new frequencies, $\omega_0+n\omega_\text{m}$, as will be seen in~\eqref{eq:omega_n}. When the periodic modulation is both spatial and temporal~\cite{Oliner_PIEEE_1963}, or spatiotemporal, as in~\eqref{eqa:S-T_perm}, we have thus the generation of an infinite number of spacetime harmonics, ($\beta_0+n\beta_\text{m},\omega_0+n\omega_\text{m}$), corresponding to oblique transitions in the dispersion diagram~\cite{Fan_NPH_2009}, as illustrated in Fig.~\ref{Fig:disp_diag}.}.In the meanwhile, the up-converted wave, $E_\text{r}=E_{1} e^{j (\omega_\text{1} t-(\beta_\text{1}-j\alpha_1) z)}$, operating in the fast wave region of the dispersion diagram, radiates as a leaky-wave under a specified angle $\theta_\text{1}$.

In the downlink, a wave with frequency $\omega_\text{1}$ is impinging on the structure under the same angle $\theta_1$ and picked up by the structure. According to phase matching, the wave can only propagate in the direction of the modulation, i.e. in the $+z$ direction, corresponding to the right-hand side of the dispersion diagram in Fig.~\ref{Fig:disp_diag}, and experiences progressive temporal frequency transition (down-conversion) from $\omega_\text{1}$ to $\omega_\text{0}=\omega_\text{1}-\omega_\text{m}$ along with spatial frequency transition from $\beta_\text{1}$ to $\beta_\text{0}=\beta_\text{1}-\beta_\text{m}$, while propagating toward the receiver port.

Given the unidirectional nature of the modulation, the structure is fundamentally nonreciprocal. No space-time transition is allowed for wave propagation in the backward ($-z$) direction (left-hand side of the dispersion diagram in Fig.~\ref{Fig:disp_diag}) due to the unavailability of modes at the points $(-\beta_0+\beta_\text{m},\omega_0+\omega_\text{m})$ from $(-\beta_0,\omega_0)$ for up-conversion and $(-\beta_1-\beta_\text{m},\omega_0)$ from $(-\beta_1,\omega_0+\omega_m)$ for down-conversion.

The direction of radiation of the main beam~\cite{Caloz_McrawHill_2011} depends on the modulation temporal and spatial frequencies as
\begin{equation}
\theta_\text{1}=\sin^{-1} \left( \frac{ \beta_1(\omega)}{ k_{01}} \right)=\sin^{-1} \left( \frac{ c(\beta_0+\beta_\text{m})}{\omega_\text{0}+\omega_\text{m}} \right),
\label{eqa:disp_unm}
\end{equation}
where $c$ is the velocity of light in vacuum and $k_{01}=\omega_1/c$ is the effective wavenumber at the frequency $\omega_1$. According to~\eqref{eqa:disp_unm}, frequency beam scanning can be achieved at a fixed input frequency, $\omega_\text{0}$, by varying the modulation frequency, $\omega_\text{m}$.
\begin{figure}
\vspace{2cm}
\centering
\psfrag{d}[c][c][0.8]{duplexer}
\psfrag{e}[c][c][0.8]{antenna, $\omega_\text{1}$}
\psfrag{B}[c][c][0.8]{LO, $\omega_\text{m}$}
\psfrag{D}[c][c][0.8]{mixer (down convertor)}
\psfrag{U}[c][c][0.8]{mixer (up convertor)}
\psfrag{F}[c][c][0.8]{\shortstack{BSF\\$\omega_\text{0}$} }
\psfrag{T}[r][c][0.8]{\shortstack{TX port,\\$\omega_\text{0}$}}
\psfrag{R}[l][c][0.8]{\shortstack{RX port,\\$\omega_\text{0}$}}
\centering{ \includegraphics[width=1\columnwidth] {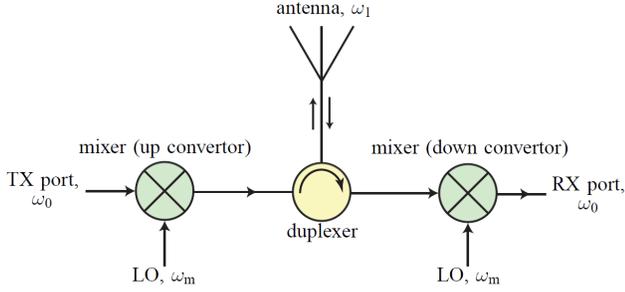}}
\vspace{-5cm}
\caption{Circuital representation of the combined mixer-duplexer-antenna operation of the space-time modulated leaky-wave structure in Fig.~\ref{Fig:ST_LW_med}.}
\label{Fig:oper_symbols}
\end{figure}

Figure~\ref{Fig:oper_symbols} shows a circuital representation of the space-time modulated leaky-wave structure of Fig.~\ref{Fig:ST_LW_med}. The operation of the structure operates as a combined mixer-duplexer-antenna system may be essentially read out from the dispersion diagram of Fig.~\ref{Fig:disp_diag}. Up-conversion and down-conversion mixing operations are provided by the oblique upwards and downwards oblique transitions, respectively. The duplexing operation (separation of uplink and downlink paths at the same frequency) is provided by nonreciprocity. Finally, the antenna operation is provided by the fast-wave nature of the wave.

\subsection{General Analytical Solution}\label{sec:gen_sol}
Since the structure is periodic in space and time, its electromagnetic field solution can be expressed as the double Floquet expansion
\begin{subequations}
\begin{equation}\label{eqa:Bloch_wave}
E(z,t)=\sum_{n =  - \infty}^\infty   E_n e^{j \omega_n t}    e^{-(\alpha_n+j \beta_n) z},
\end{equation}
where $E_n$ is the amplitude of the $n^\text{th}$ space-time harmonic, characterized by the temporal and spatial frequencies
\begin{equation}\label{eq:omega_n}
\omega_n=\omega_\text{0}+n\omega_\text{m}
\end{equation}
and
\begin{equation}\label{eq:beta_n}
\beta_n(\omega)=\beta_0(\omega)+n\beta_\text{m},
\end{equation}
\end{subequations}
respectively, and where $\alpha_n(\omega)$ is the leakage factor of the $n^\text{th}$ harmonic. For a given input amplitude $E_0$ and frequency $\omega_0$, $E_n$ in~\eqref{eqa:Bloch_wave}, leading to the general field solution, and $\beta_0(\omega)$ in~\eqref{eq:beta_n}, corresponding to the dispersion relation, are unknowns to be determined. The detailed resolution of this problem is presented in Appendix~\ref{App_gen}.

In the case of weak modulation, $\delta_\text{m}\ll 1$, the $E_n$'s in~\eqref{eqa:Bloch_wave} for $|n|>1$ are negligible. Moreover, assuming that the leaky-wave structure is designed to support uplink and downlink radiation mainly at $\omega _\text{1}=\omega _\text{0}+\omega _\text{m}$, in the uplink, the amplitude of the lower harmonic $E_{-1}$, corresponding to the frequency $\omega _\text{0}-\omega _\text{m}$, and in the downlink, the amplitude of the higher harmonic $E_{2}$ (already negligible from the weak modulation assumption), corresponding to the frequency $\omega _\text{1}+\omega _\text{m}=\omega _\text{0}+2\omega _\text{m}$, are negligible. Then we find, as derived in Appendix~\ref{App_up_down_conv}, assuming $\alpha_n\ll \beta_n$, that
\begin{equation}\label{eqa:up_disp_rel_body}
\beta_\text{0}(\omega)=\pm \beta_\text{um}(\omega_0) \pm \frac{\delta_\text{m}}{4} \sqrt{\beta_\text{um}(\omega_0) \beta_\text{um}'(\omega_0) },
\end{equation}
where the upper and lower signs correspond to forward and backward propagation, respectively, and where ${\beta_\text{um}'(\omega_0) =\beta_\text{um}(\omega_0)+\beta_\text{m}}$ with $\beta_\text{um}$ being the wavenumber of the unmodulated structure, which, according to~\eqref{eqa:disp_unm} with $\beta_\text{m}=\omega_\text{m}=0$, may be obtained from the main beam radiation angle as

\begin{equation}\label{eqa:beta_unmod}
\beta_\text{um}(\omega_0)
=\frac{\omega_\text{0}}{c}\sin(\theta(\omega_0)).
\end{equation}
Equation~\eqref{eqa:up_disp_rel} shows that, as expected, the change in the wave vector due to the modulation is proportional to the modulation depth, $\delta_\text{m}$, and to the modulation wavenumber, $\beta_\text{m}$.

As shown in Appendix~\ref{App_up_down_conv}, the amplitude of the uplink up-converted electric field is
\begin{equation}
E_{1} =  \frac{\delta_\text{m} E_0 \beta_\text{um}'}{\delta_\text{m} \sqrt{\beta_\text{um} \beta_\text{um}'}-2\alpha_1^2/\beta_\text{um}'-j \alpha_1\left(\delta_\text{m}  \sqrt{\beta_\text{um}/\beta_\text{um}'}+4 \right)},
\label{eqa:E_up}
\end{equation}
which is proportional to the input electric field, $E_0$, and depends on both the modulation depth, $\delta_\text{m}$, and the leakage factor, $\alpha_1$. As shown in~\eqref{eqa:trans_conv}, uplink conversion is associated with power conversion gain.

Similarly, as shown in Appendix~\ref{App_up_down_conv}, the amplitude of the downlink down-converted electric field, corresponding to $n=0$, is
\begin{equation}
E_{0} =  \frac{\delta_\text{m} E_1 \beta_\text{um}}{\delta_\text{m} \sqrt{\beta_\text{um} \beta_\text{um}'}-2\alpha_0^2/\beta_\text{um}-j \alpha_0\left(\delta_\text{m}  \sqrt{\beta_\text{um}'/\beta_\text{um}}+4 \right)},
\label{eqa:E_down}
\end{equation}
which is proportional to the received electric field, $E_1$, and depends on both the modulation depth, $\delta_\text{m}$, and the leakage factor, $\alpha_0$, and which is associated with power conversion loss.

\section{Mixer-Duplexer-Antenna System Realization}\label{sec:Mix_Dupl_LWA}

The space-time modulated permittivity in~\eqref{eqa:S-T_perm} is realized as follows. An array of sub-wavelengthly spaced varactors is distributed in parallel with the intrinsic capacitance of a microstrip transmission line, while a sinusoidal modulation signal, with frequency $\omega_\text{m}$ and wave vector $\beta_\text{m}$, propagating in the guided-mode regime along $+z$ direction, modulates the varactors in space and time such as to provide the space-time dependent capacitance $C(z,t)=C_\text{e}+C_\text{var} \cos(\omega_\text{m}t-\beta_\text{m}z)$. As a result, the corresponding effective permittivity of the medium becomes $\epsilon(z,t)= \epsilon_\text{e}(1+\delta_\text{m} \cos(\omega_\text{m}t-\beta_\text{m}z))$, where the modulation depth, $\delta_\text{m}$, depends on the range of variation of the varactors at a given modulation amplitude. Note that a DC bias line is used to set the varactors in the reverse bias (capacitive) regime.

\begin{figure}
\vspace{4.9cm}
\centering{ \includegraphics[width=2.4\columnwidth]{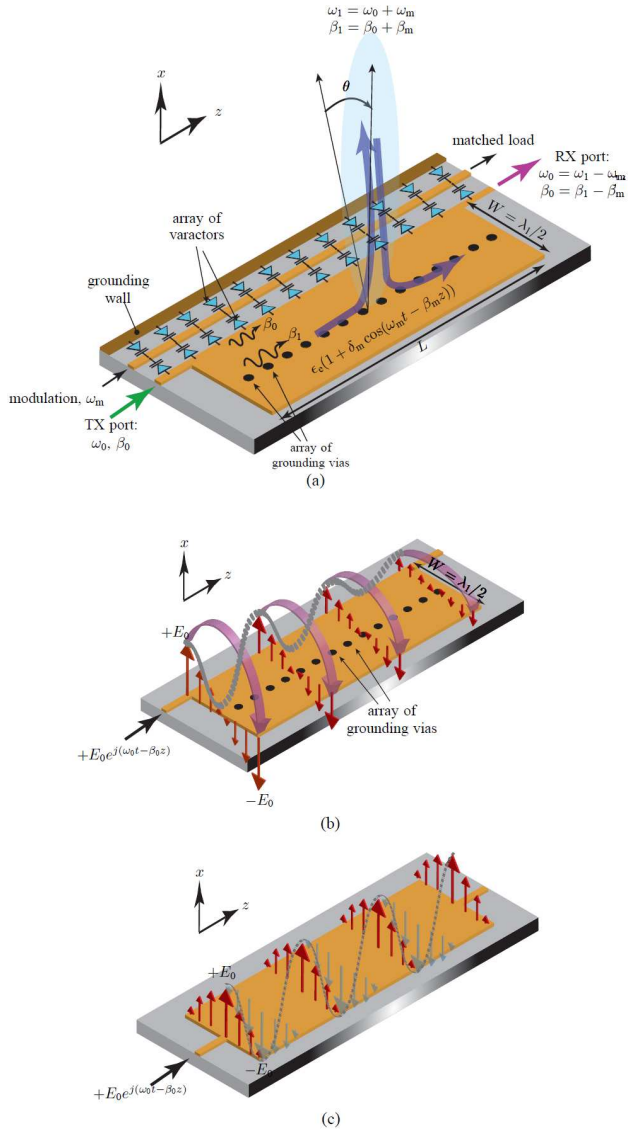}}
\vspace{-6.3cm}
\caption{Realization of the space-time modulated mixer-duplexer-antenna system in Fig.~\ref{Fig:oper_symbols}. (a)~Structure, based on the half-wavelength microstrip leaky-wave antenna~\cite{Menzel_EuMC_1978}. (b)~Leaky-mode ($EH_1$) odd electric field distribution of along the structure, unaffected by the vias given their position in the nodal plane of the mode ($xz$ plane, middle of the strip). (c)~Dominant guided-mode (quasi-TEM) even electric field distribution with antinode in the plane mentioned in (b) and hence shorting (suppression) of this mode in the presence of the vias.}
\label{Fig:str_details}
\end{figure}

Figure~\ref{Fig:str_details}(a) shows the structure and operation of the realized space-time modulated mixer-duplexer-antenna system. The structure supports two modes, a dominant even mode, providing a guided-mode channel for the wave modulating the varactors (narrow strip in Fig.~\ref{Fig:str_details}) and a higher order odd ($EH_1$) mode, providing the leaky-mode channel for radiation (wide strip in Fig.~\ref{Fig:str_details}). The transmit and receive ports are designed in such a way as to excite the leaky mode of the microstrip line. Moreover, an array of grounding vias is placed at the center of the wide microstrip line to suppress the even mode and hence enforce optimal leaky-wave radiation at $\omega_\text{1}$, which corresponds to a strip width of $W=\lambda_\text{1}/2$. Figure~\ref{Fig:str_details}(b) shows the electric field distribution of the leaky-mode ($EH_1)$ along the microstrip leaky-wave antenna~\cite{Menzel_EuMC_1978,Oliner_McrawHill_2007,Caloz_McrawHill_2011}. In contrast, Fig.~\ref{Fig:str_details}(c) shows the electric field distribution of the dominant guided-mode ($EH_0)$ of a microstrip transmission line, where transversally symmetric distribution of the electric field yields a complete propagation of the wave. The narrow line in the realized system of Fig.~\ref{Fig:str_details} supporting the propagation of modulation signal is designed in such a way to support the propagation of the dominant guided-mode of the microstrip line.

\section{System Implementation and Characterization}\label{sec:prototype}

This section presents the implementation and characterization of the mixer-duplexer-antenna system. The modulation specifications are $f_\text{m}=\omega_\text{m}/2\pi=0.18$~GHz, $\beta_\text{m}=\omega_\text{m}\sqrt{\epsilon_\text{e}}/c=5.16$~rad/m, $\delta_\text{m}=0.15$. The modulation circuit is composed of 39 unit cells of antiparallel varactors [Fig.~\ref{Fig:str_details}], with uniform spacing, or period, of $p=5$~mm, which corresponds to $p/\lambda_\text{m}=p\beta_\text{m}/(2\pi)\approx 1/250$, and hence safely satisfies medium homogeneity in accordance with~\eqref{eqa:S-T_perm}. We employed BB833 varactors manufactured by Infineon Technologies where the capacitance ratio (highest over lowest capacitance) is about 12.4. The specifications of the structure are $f_\text{0}=1.7$~GHz, $f_\text{1}=1.88$~GHz, $\theta_1=4^\circ$ (radiation angle corresponding to the frequency $f_1$), $\alpha_0=1.2$~Np/m, $\alpha_1=3.4$~Np/m, $L=8$~in and $W=1.8$~in, RT5880 substrate with permittivity $\epsilon_\text{r} = 2.2$, thickness $h = 125$~mil and $\tan \delta = 0.0009$\footnote{The leakage factors, $\alpha_0$ and $\alpha_1$, are experimentally obtained from the scattering parameter $S_{21}$ as $\alpha_k\approx-\ln|S_{21}(\omega_k)|/L$, $k=0,1$, since most of the attenuation is due to leakage.}.

\begin{figure}
\centering
\psfrag{A}[c][c][0.85]{$\beta L$}
\psfrag{B}[c][c][0.85]{$\omega/2\pi~\text{(GHz)}$}
\psfrag{C}[l][c][0.8]{($f_\text{0},\beta_\text{0}$)}
\psfrag{D}[l][c][0.8]{($f_\text{1},\beta_\text{1}$)}
\psfrag{E}[l][c][0.8]{unmodulated case, $\beta_\text{um}(\omega)$}
\psfrag{F}[l][c][0.8]{modulated case, $\beta(\omega)$}
\psfrag{G}[c][c][0.8]{air-line}
\psfrag{H}[c][c][0.8]{\shortstack{guided-\\mode}}
\centering{ \includegraphics[width=0.9\columnwidth] {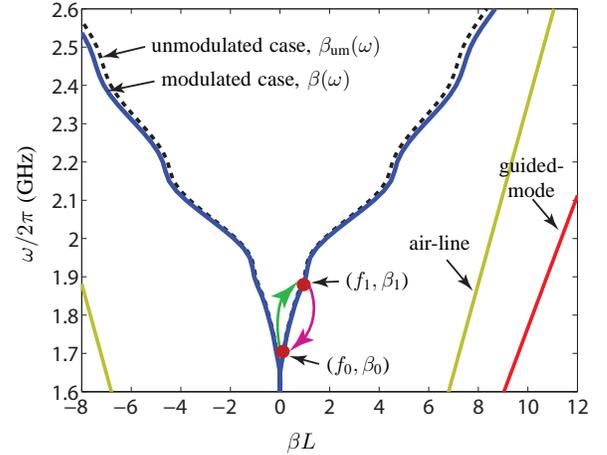}}
\caption{Dispersion diagram for the prototype depicted in Fig.~\ref{Fig:str_details} computed using~\eqref{eqa:dispers_relat} and~\eqref{eqa:beta_unmod}.}
\label{Fig:disp_anal}
\end{figure}

\begin{figure}
\vspace{2cm}
\centering
\psfrag{A}[c][c][0.85]{$\beta_\text{um} L$}
\psfrag{B}[c][c][0.85][90]{Error ($\%$)}
\psfrag{C}[c][c][0.85]{$\delta_\text{m}=0.05$}
\psfrag{D}[c][c][0.85]{$\delta_\text{m}=0.15$}
\centering{\includegraphics[width=0.9\columnwidth] {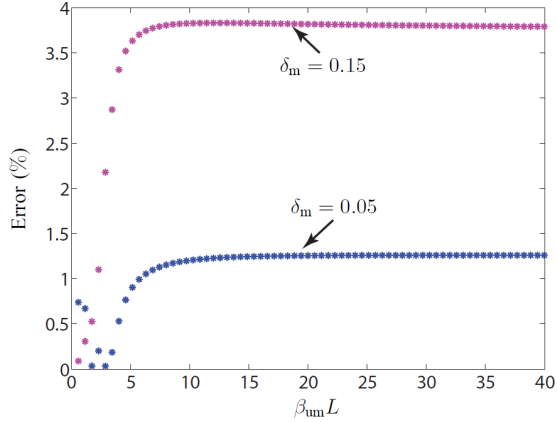}}
\vspace{-2.4cm}
\caption{Relative error, ($\mid1-\beta(\omega)/\beta^\text{exact}(\omega)\mid\times 100$), of the approximate dispersion, $\beta(\omega)$, computed by~\eqref{eqa:up_disp_rel_body}, with respect to the exact dispersion, $\beta^\text{exact}(\omega)$, numerically extracted from~\eqref{eqa:dispers_relat} for two specific modulation depths.}
\label{Fig:error_aprox_beta}
\end{figure}

\begin{figure}
\vspace{3.5cm}
\begin{center}
\includegraphics[width=1.9\columnwidth]{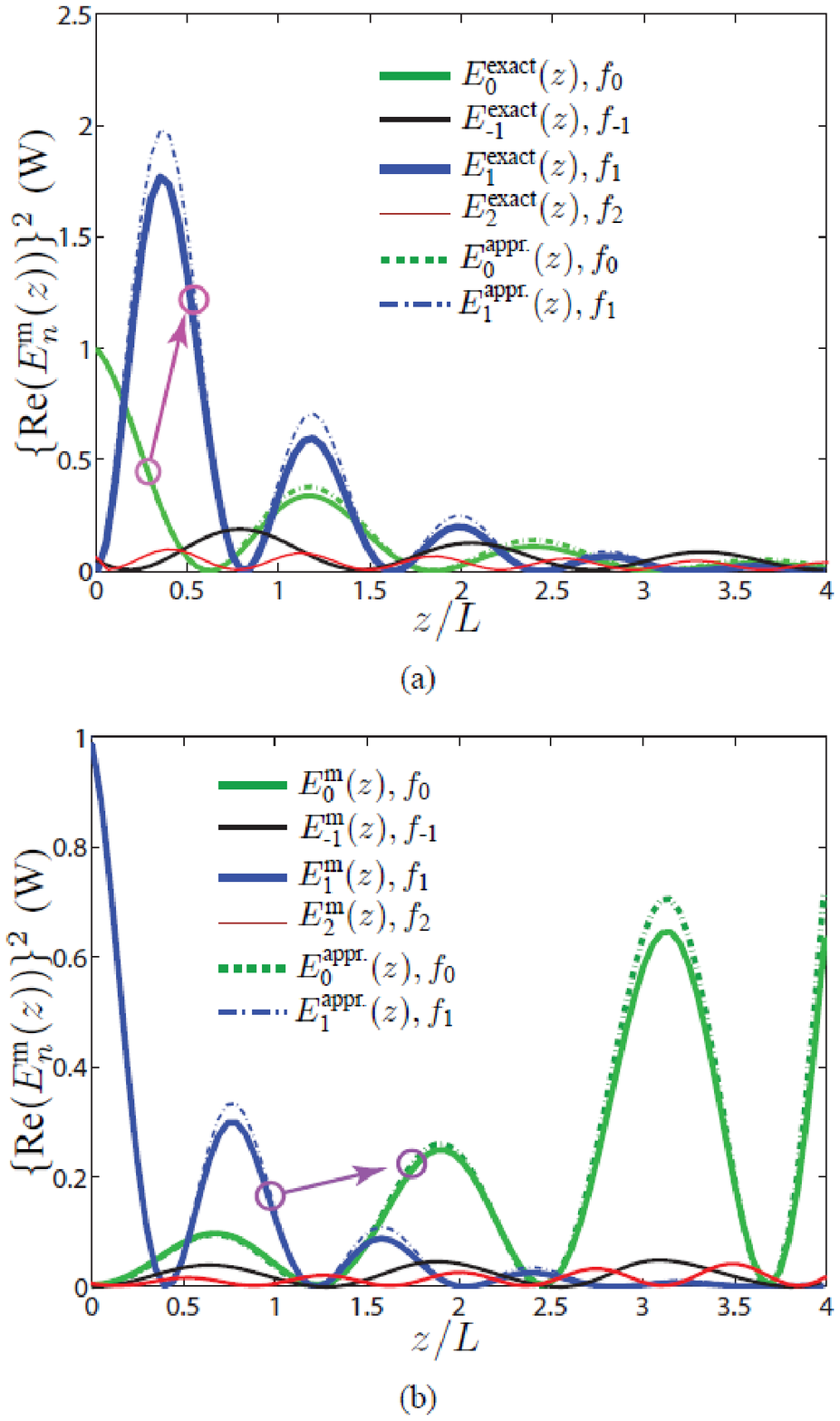}
\vspace{-5.3cm}
\caption{Squared real part of the electric field space-time harmonics, propagating in the $+z$ direction along the antenna for the (a)~uplink and (b)~downlink, where the approximate wave solutions for the up- and down-links, $E_\text{0}^\text{appr.}(z)$ and $E_\text{1}^\text{appr.}(z)$, obtained using~\eqref{eqa:E_up} and~\eqref{eqa:E_down} are compared with the exact wave solutions obtained using~\eqref{eqa:Bloch_wave} with~\eqref{eqa:E_neg_bn} and~\eqref{eqa:E_pos_bn}.}
\end{center}
\label{Fig:field_amp}
\end{figure}

Figure~\ref{Fig:disp_anal} shows the dispersion diagram of the realized prototype computed using~\eqref{eqa:dispers_relat} and~\eqref{eqa:beta_unmod} with varying $\omega_0$. This diagram is, as expected, qualitatively similar to that in Fig.~\ref{Fig:disp_diag}, with a leaky-mode cutoff frequency of about 1.65~GHz. It may be observed that the dispersion of the modulated mode corresponds, also as expected, to an increased momentum ($\beta$), although the difference with the unmodulated dispersion is negligible in the transition range, consistently with the weak modulation assumption. In uplink (green arrow), the wave is up-converted from $f_\text{0}=1.7$~GHz to $f_\text{1}=1.88$~GHz while in the downlink (blue arrow), it is down-converted from $f_\text{1}=1.88$~GHz to $f_\text{0}=1.7$~GHz\footnote{The ratio between the frequency pair ($f_1,f_0$) is practically limited in terms the maximal acceptable size of the antenna. Indeed, as seen in Fig.~\ref{Fig:field_amp}, the antenna must be sufficiently long for the input power at $f_0$ to sufficiently convert, for a specified conversion efficiency, to the output power at $f_1$, and the antenna must have a length that is at least a couple of wavelengths of the \emph{lowest} frequency. This means that, if $f_1/f_0=\kappa$, then the antenna must be $\kappa$ times longer than an antenna that would be conventionally operated at $f_1$.}. Figure~\ref{Fig:error_aprox_beta} shows the relative error of the approximate dispersion relation, given by~\eqref{eqa:up_disp_rel_body}, with respect to the exact dispersion relation, obtained by~\eqref{eqa:dispers_relat}, for the modulation indices $\delta_\text{m}=$ $0.05$ and $0.15$. As expected, the error is proportional to the modulation depth. Moreover, we see that the approximate solution of the dispersion diagram for $\delta_\text{m}=0.15$, corresponding to the design of Fig.~\ref{Fig:disp_anal} would be of less than $4\%$.

Figure 7(a) shows the exact and approximate squared real part of the field propagating along the structure for the uplink, computed using~\eqref{eqa:Bloch_wave} with~\eqref{eqa:E_neg_bn} and~\eqref{eqa:E_pos_bn} for the exact solutions and using~\eqref{eqa:E_up} and~\eqref{eqa:E_down} for the approximate solutions. The power of the input wave ($E_0^\text{m}$, $f_0=1.7$~GHz) progressively transfers to the leaky wave ($E_1^\text{m}$, $f_0=1.88$~GHz) along the $+z$ direction, with conversion gain, as predicted in Sec.~\ref{sec:gen_sol}. At the same time, the power of the leaky mode exponentially decreases due to radiation. Similarly, Fig. 7(b) shows the squared real part of the field propagating along the structure for the downlink, computed using~\eqref{eqa:E_down}. Here, the power of the incoming wave ($E_1^\text{m}$, $f_0=1.88$~GHz) progressively transfers, still along the $+z$ direction, to the output wave ($E_0^\text{m}$, $f_0=1.7$~GHz). We see in Figs. 7(a) and 7(b) that the power level and, more importantly, the power decay, which corresponds to leaky-wave radiation, is much less for higher order space-time harmonics, i.e. $E_{-1}^\text{m}$ and $E_2^\text{m}$, than it is for the input and radiating harmonics, $E_{0}^\text{m}$ and $E_1^\text{m}$.

\begin{figure}
\centering
\psfrag{A}[c][c][0.7]{DC-block}
\psfrag{B}[c][c][0.7]{array of varactors}
\psfrag{F}[c][c][0.7]{\shortstack{DC-input\\varactors bias}}
\psfrag{D}[c][c][0.7]{\shortstack{matched load\\(mod. output)}}
\psfrag{G}[c][c][0.7]{\shortstack{array of\\grounding vias}}
\psfrag{E}[c][c][0.7]{$f_\text{0}$}
\psfrag{L}[c][c][0.7]{$L$}
\psfrag{C}[c][c][0.7]{$C$}
\psfrag{T}[l][c][0.7]{TX}
\psfrag{R}[l][c][0.7]{RX}
\psfrag{M}[l][c][0.7]{mod.}
\psfrag{E}[c][c][0.7]{\shortstack{band-stop filter\\at $f_\text{0}=1.7~$GHz\\(tunable)}}
\centering{ \includegraphics[width=0.95\columnwidth] {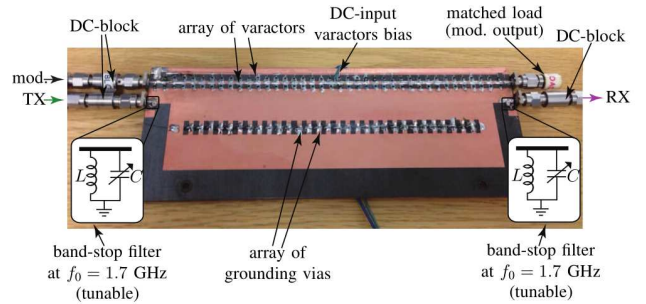}}
\caption{Photograph of the fabricated prototype.}
\label{Fig:photo_LWA}
\end{figure}

Figure~\ref{Fig:photo_LWA} shows a photograph of the realized prototype. Band-stop filters centered at $f_0$ are used at the transmit and receive ports to ensure the suppression of unwanted harmonics, such as for instance $f_\text{m}$ and $f_{-1}$.

\section{Full-wave and Experimental Results}\label{sec:exp_res}

\subsection{Matching and Measurement Setup}

\begin{figure}[h]
\centering
\psfrag{A}[c][c][0.9]{Frequency (GHz)}
\psfrag{B}[c][c][0.9]{TX and RX ports matching (dB)}
\psfrag{C}[l][c][0.9]{full-wave}
\psfrag{D}[l][c][0.9]{measurement}
\psfrag{E}[c][c][0.9]{$f_\text{0}$}
\centering{ \includegraphics[width=0.75\columnwidth] {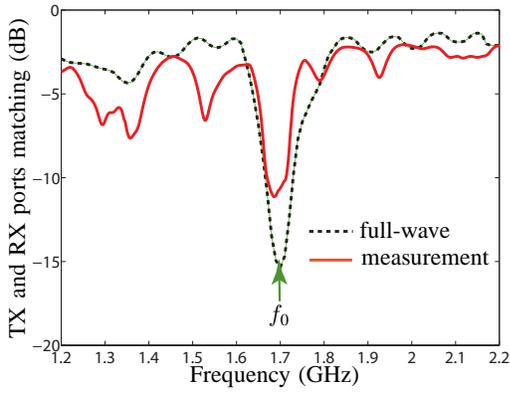}}
\caption{Matching of the transmit and receive ports of the antenna system in Fig~\ref{Fig:photo_LWA}. }
\label{Fig:s11}
\end{figure}

Figure~\ref{Fig:s11} shows the full-wave and experimental matching of the transmit and receive ports of the antenna system in Fig~\ref{Fig:photo_LWA} for an optimal varactor DC bias of 12~V\footnote{At this voltage level, the component safely operates in the \emph{linear} reverse-biased regime, so that related nonlinear effects are negligible.} and the band-stop filters at both ports tuned at $f_0=1.7$~GHz.

\begin{figure}
\vspace{2.2cm}
\centering
\psfrag{a}[l][c][0.9]{\shortstack{TX, $f_\text{0}$\\Port-1}}
\psfrag{b}[c][c][0.9]{\shortstack{reference antenna, $f_\text{1}$\\Port-3}}
\psfrag{c}[l][c][0.9]{\shortstack{RX, $f_\text{0}$\\Port-2}}
\psfrag{d}[l][c][0.9]{\shortstack{$P_{13}$}}
\psfrag{e}[l][c][0.9]{\shortstack{$P_{21}$}}
\psfrag{f}[l][c][0.9]{\shortstack{$P_{12}$}}
\psfrag{g}[l][c][0.9]{\shortstack{$P_{23}$}}
\psfrag{h}[l][c][0.9]{\shortstack{$P_{32}$}}
\centering{ \includegraphics[width=0.9\columnwidth]{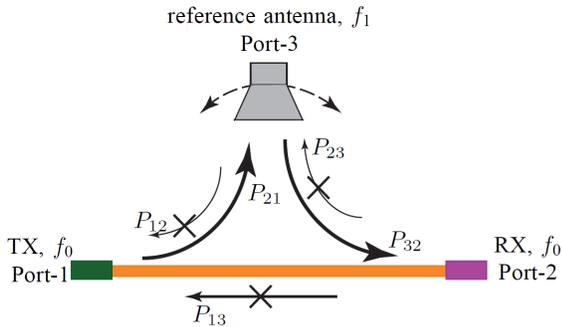}}
\vspace{-3.8cm}
\caption{Experimental setup for the measurement of the mixer-duplexer-antenna system.}
\label{Fig:oper}
\end{figure}

\begin{figure}
\vspace{3.2cm}
\centering
\includegraphics[width=1.7\columnwidth]{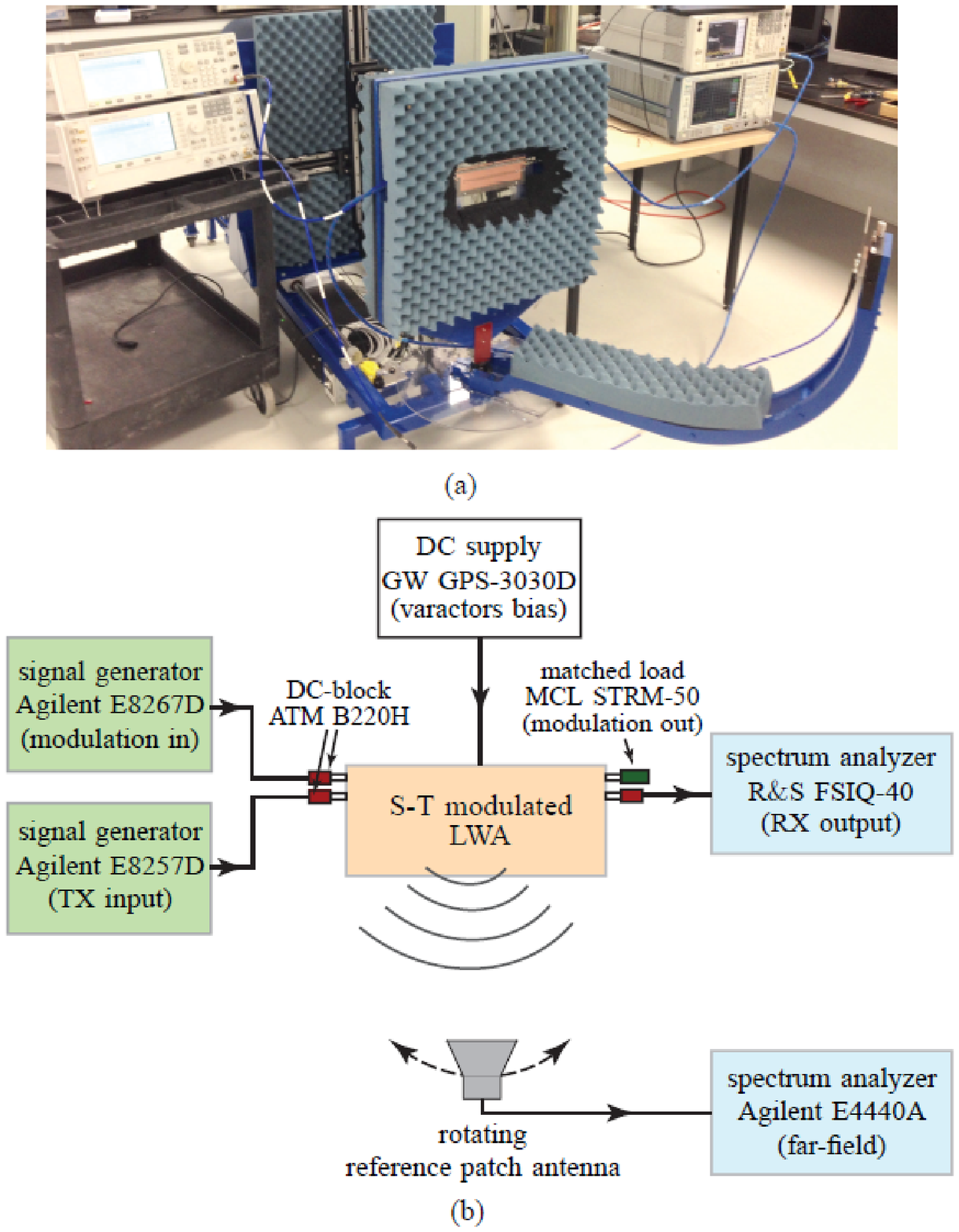}
\vspace{-4.5cm}
\caption{Measurement set-up. (a) Photograph. (b) Schematic.}
\label{Fig:meas_setup}
\end{figure}

Figure~\ref{Fig:oper} shows the experimental setup, which uses a rotating reference antenna as a third port to model far-field radiation in both the transmit and receive regimes. Figure~\ref{Fig:meas_setup}(a) and~(b) show the photograph and schematic of the complete measurement setup, respectively. An Agilent E8267D signal generator, set at frequency $f_\text{m}=0.18$~GHz and amplitude $P_\text{m}=15$~dBm, provides the modulation signal for the varactors. The distance between the reference and test antennas is about 1~m (6.3$\lambda_0$ at the radiation frequency $f_1=1.88$~GHz), which is beyond the far-field distance, $r_\text{far-field}\approx$~0.5~m. For the uplink, an Agilent E825D signal generator provides the input signal at the transmit port, with frequency $f_\text{0}=1.7$~GHz and amplitude $P_\text{0}=0$~dBm, while two spectrum analyzers (R$\&$S FSIQ-40 and Agilent E4440A) measure the received power at the receive port and at the port of the rotating reference antenna. In the downlink, the Agilent E825D signal generator provides an input signal at the rotating reference antenna port, with frequency $f_\text{1}=1.88$~GHz and amplitude $P_\text{0}=0$~dBm, while the two spectrum analyzers measure the received power at the receive and transmit ports.


\subsection{Fixed Radiation Beam}

\begin{figure}
\vspace{2.2cm}
\begin{center}
\includegraphics[width=1.1\columnwidth]{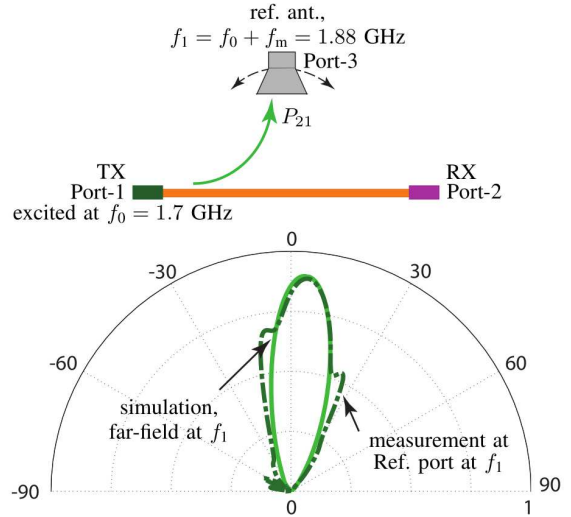}
\vspace{-3.4cm}
\caption{Uplink normalized power full-wave and measurement results.}
\label{Fig:result_trans}
\end{center}
\end{figure}

\begin{figure}
\begin{center}
\subfigure[]{\label{Fig:result_rec-a}
\psfrag{A}[r][c][0.8]{\shortstack{$f_\text{1}=1.88$}}
\psfrag{B}[l][c][0.8]{1.92}
\psfrag{C}[l][c][0.8]{1.97}
\psfrag{D}[l][c][0.8]{2~GHz}
\psfrag{a}[c][c][0.8]{\shortstack{TX\\Port-1}}
\psfrag{b}[c][c][0.8]{\shortstack{ref. ant.\\excited at $f_\text{1}=1.88$~GHz}}
\psfrag{P}[l][c][0.8]{Port-3}
\psfrag{c}[l][c][0.8]{\shortstack{RX\\Port-2}}
\psfrag{h}[l][c][0.8]{\shortstack{$P_{32}$}}
\psfrag{H}[l][c][0.8]{\shortstack{$P_{12}$}}
\psfrag{M}[c][c][0.8]{$f_\text{0}=f_\text{1}-f_\text{m}=1.7$~GHz}
\psfrag{W}[c][c][0.9]{\shortstack{measurement at\\TX port at $f_\text{0}$}}
\psfrag{Z}[c][c][0.9]{\shortstack{measurement at\\RX port at $f_\text{0}$}}
\includegraphics[width=0.85\columnwidth]{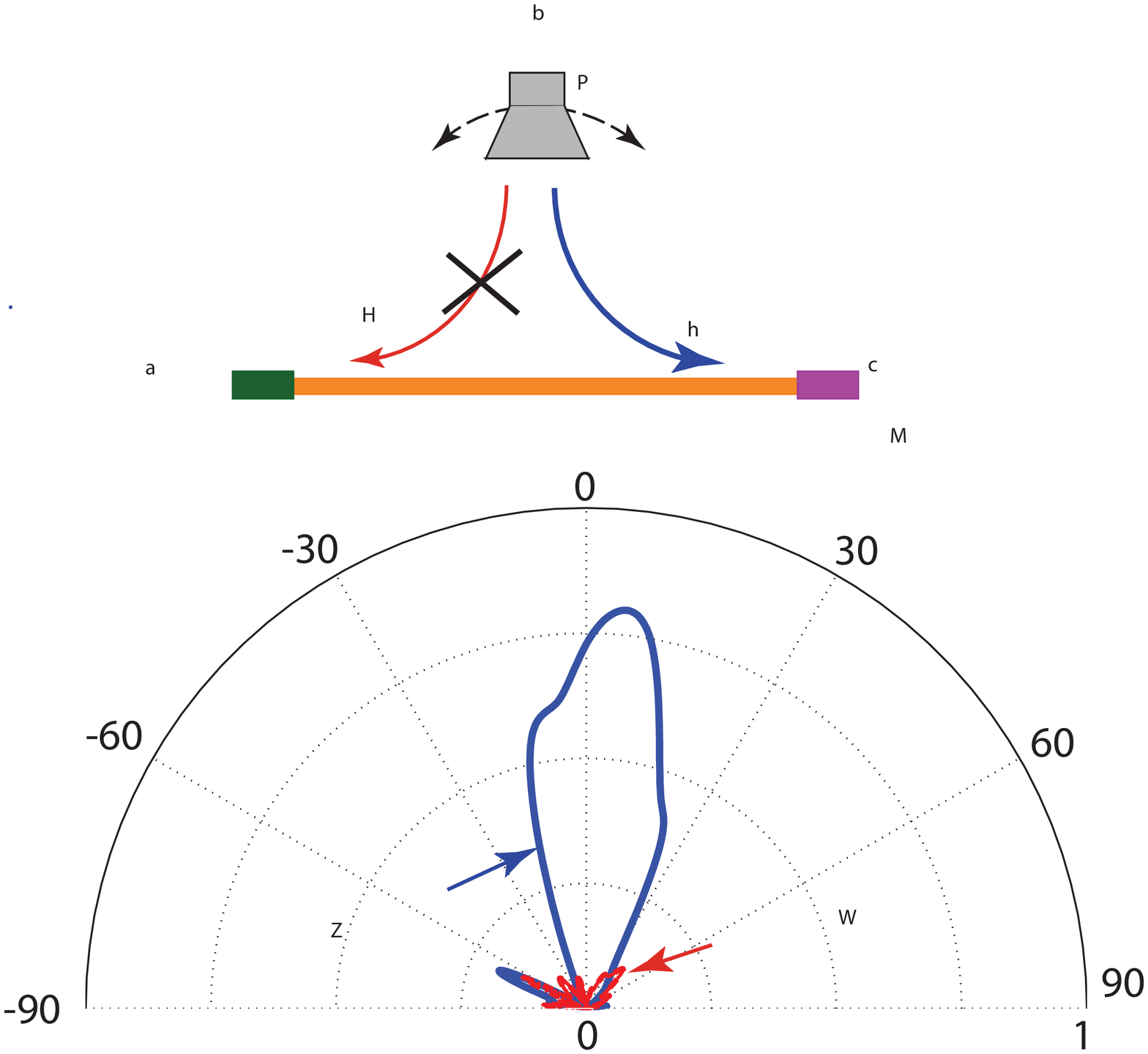}}
\subfigure[]{\label{Fig:result_rec-b}
\psfrag{A}[c][c][0.8]{$\theta$ Degree}
\psfrag{B}[c][c][0.8]{\shortstack{Isolation between RX and TX ports\\in receive mode (dB)}}
\psfrag{C}[c][c][0.8]{\shortstack{radiation\\angle}}
\includegraphics[width=0.85\columnwidth]{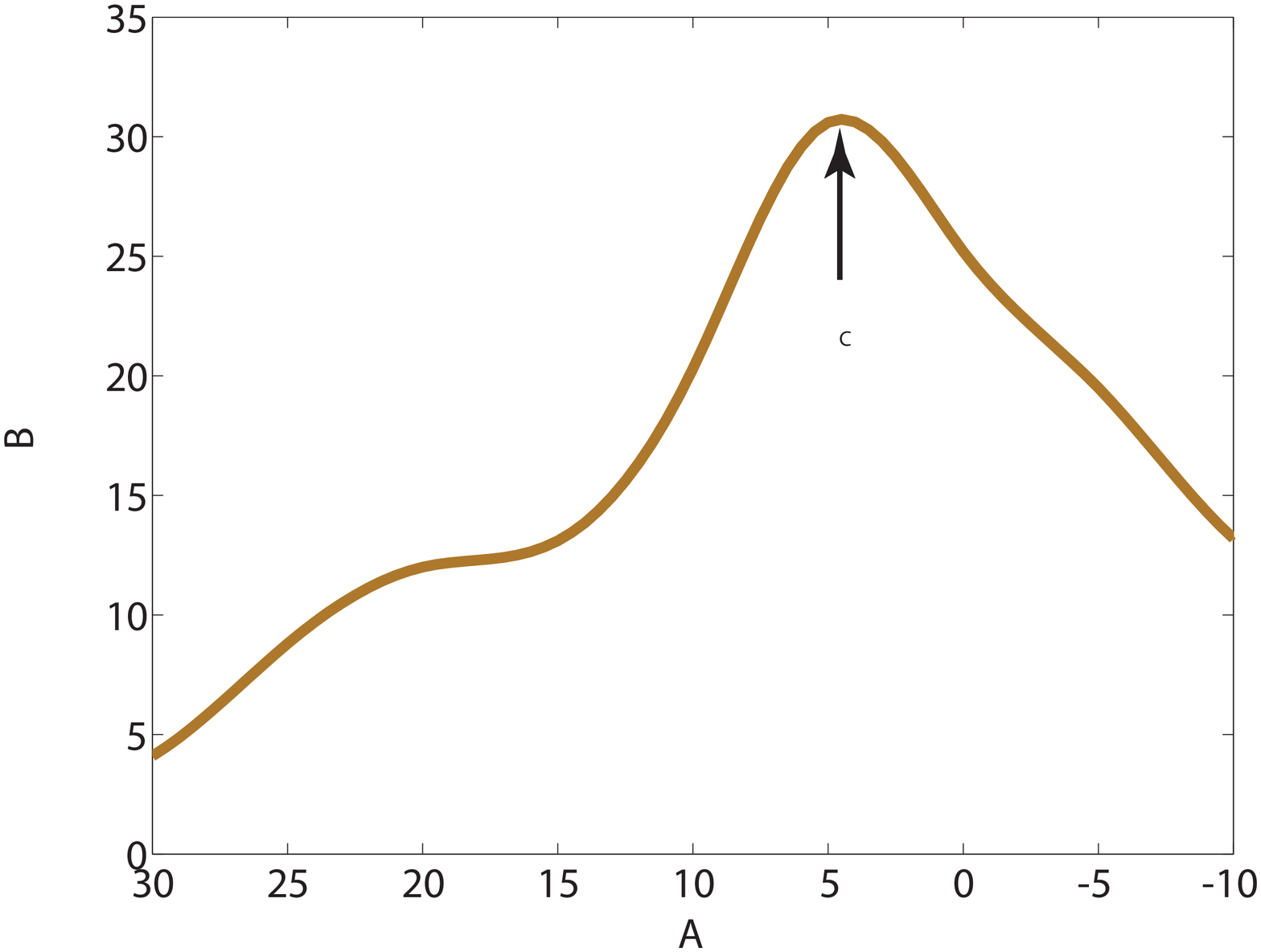}}
\caption{Downlink measurement results. (a)~Normalized received power at the receive and transmit ports at $f_\text{0}=1.7$~GHz for the incoming wave with frequency $f_\text{1}=1.88$~GHz. (b)~Isolation between received power at the receive and transmit ports.}
\end{center}
\end{figure}

Figure~\ref{Fig:result_trans} shows the full-wave and experimental normalized radiated powers at $f_\text{1}=$1.88~GHz for the uplink, when the transmit port is excited at $f_\text{0}=1.7$~GHz. The maximum of the radiated power is at $\theta_1=4^\circ$. Figure~\ref{Fig:result_rec-a} shows the full-wave and experimental normalized received powers at the receive and transmit ports for the downlink at $f_\text{0}=1.7$~GHz, when the reference antenna port is excited at $f_\text{1}=1.88$~GHz. The radiation efficiency is 74~$\%$; it is mainly limited by the shortness of the antenna (due to fabrication limitation), as understood from Fig.~\ref{Fig:field_amp}. Figure~\ref{Fig:result_rec-b} plots the isolation between the powers received at the receive and transmit ports. The isolation achieved at specified radiation angle ($\theta_1=4^\circ$) is about 31.5~dB. The discrepancy between the simulation and measurement results are mainly attributed to the imperfect modeling of the varactors in the simulation.
\subsection{Frequency Beam Scanning}

Frequency beam scanning is one of the interesting properties of leaky-wave antennas, where the radiation beam angle can be controlled by the operation frequency of the antenna~\cite{Caloz_McrawHill_2011}. However, it is more practical to control the radiation beam angle at fixed input frequency. For this reason, we perform here frequency beam scanning at fixed input frequency ($f_\text{0}$) by varying the modulation frequency ($f_\text{m}$) since this also results in varying the radiation frequency ($f_\text{1}=f_\text{0}+f_\text{m}$).

\begin{figure}
\vspace{3.6cm}
\begin{center}
\includegraphics[width=2.1\columnwidth]{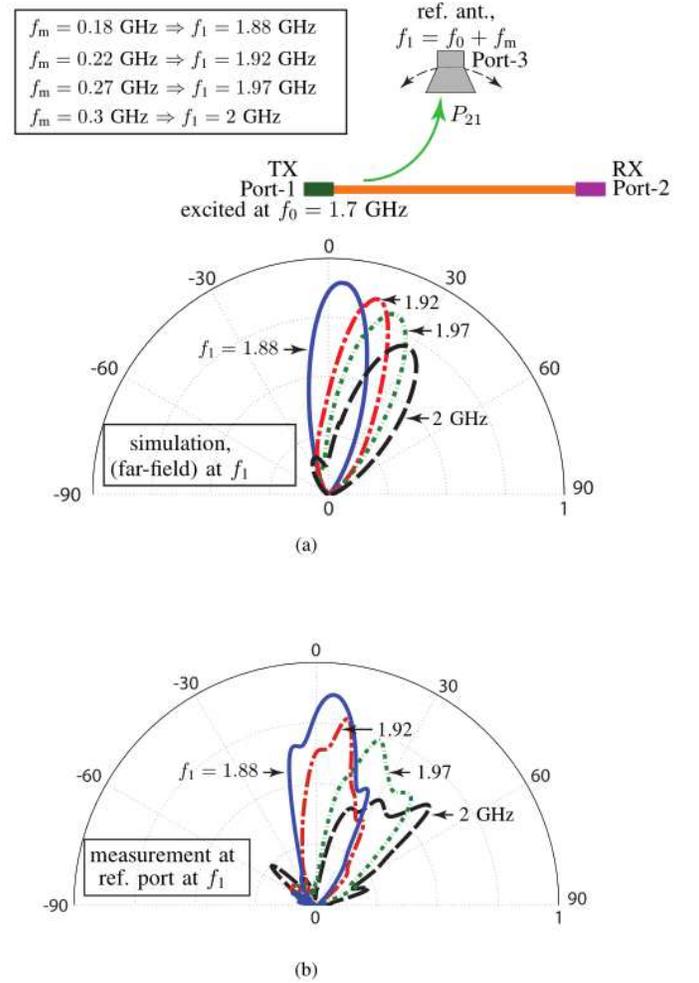}
\vspace{-5.5cm}
\caption{Uplink normalized scanning radiated power at $\theta_1=\{4, 11.5, 18, 24.5\}^\circ$ for $f_\text{1}=\{1.88, 1.92, 1.97, 2\}$~GHz corresponding to $f_\text{m}=\{0.18, 0.22, 0.27, 0.3\}$~GHz for the input frequency $f_\text{0}=1.7$~GHz. (a)~Simulation results. (b)~Measurement.}
\end{center}\label{Fig:res_trans_scanning-a}
\end{figure}

Figures~\ref{Fig:res_trans_scanning-a}(a) and~\ref{Fig:res_trans_scanning-a}(b) show the full-wave and experimental normalized radiated powers for uplink frequency beam scanning. The input frequency is $f_\text{0}=1.7$~GHz, and varying the modulation frequency as $f_\text{m}=\{0.18, 0.22, 0.27, 0.3\}$~GHz yields the radiation frequencies $f_\text{1}=\{1.88, 1.92, 1.97, 2\}$~GHz, corresponding to the radiation beam angles of $\theta_1=\{4, 11.5, 18, 24.5\}^\circ$.

\begin{figure}
\begin{center}
\subfigure[]{\label{Fig:res_rec_scanning-a}
\psfrag{A}[r][c][0.8]{\shortstack{$f_\text{1}=1.88$}}
\psfrag{B}[l][c][0.8]{1.92}
\psfrag{C}[l][c][0.8]{1.97}
\psfrag{D}[l][c][0.8]{2~GHz}
\psfrag{a}[c][c][0.8]{\shortstack{TX\\Port-1}}
\psfrag{b}[c][c][0.8]{\shortstack{ref. ant.\\excited at $f_\text{1}$}}
\psfrag{P}[l][c][0.8]{Port-3}
\psfrag{c}[l][c][0.8]{\shortstack{RX\\Port-2}}
\psfrag{h}[l][c][0.8]{\shortstack{$P_{32}$}}
\psfrag{M}[c][c][0.8]{$f_\text{0}=f_\text{1}-f_\text{m}=1.7$~GHz}
\psfrag{W}[l][l][0.9]{\shortstack{measurement at\\RX port at $f_\text{0}$}}
\includegraphics[width=0.9\columnwidth]{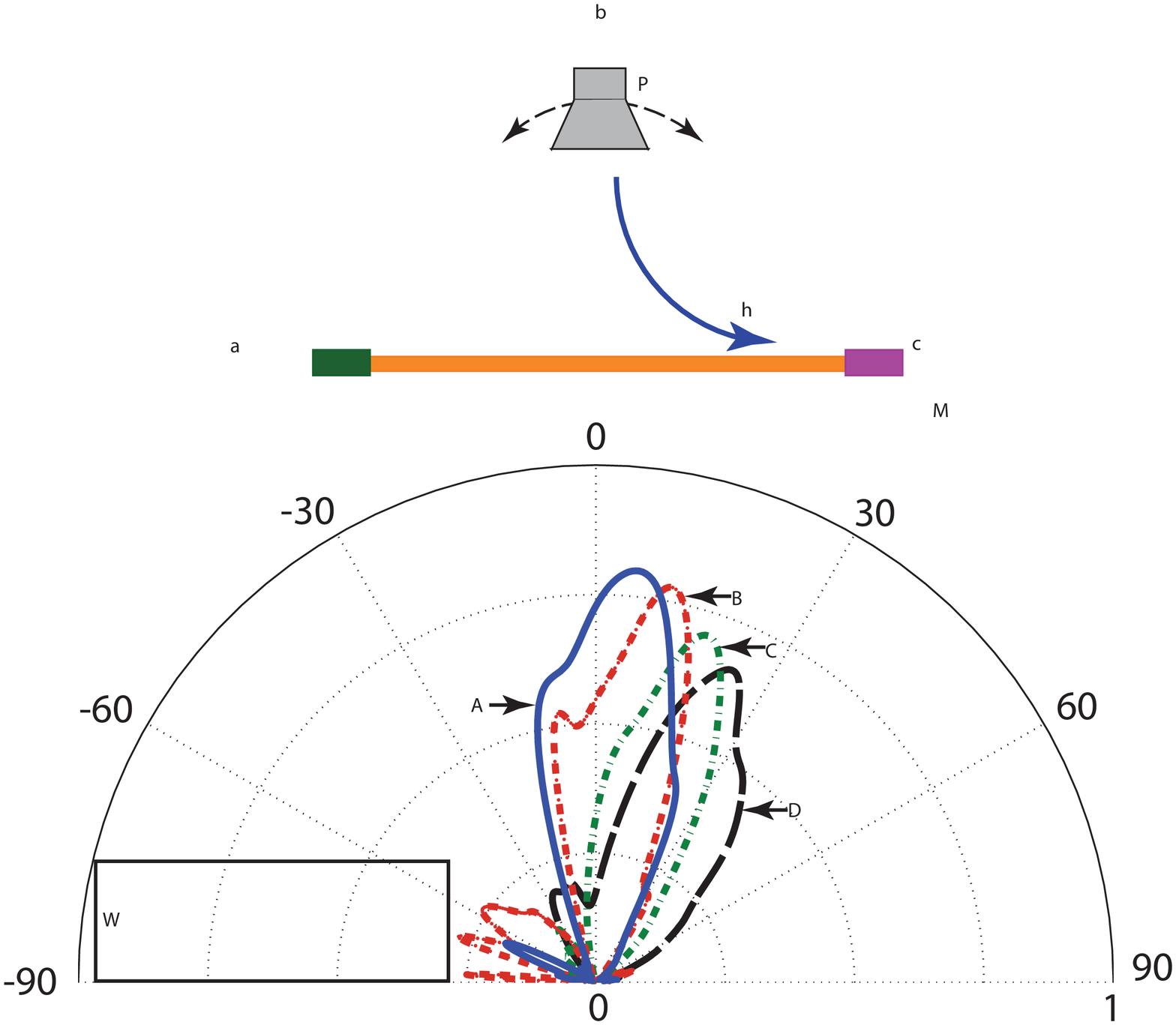}}
\subfigure[]{\label{Fig:res_rec_scanning-b}
\psfrag{a}[c][c][0.8]{\shortstack{TX\\Port-1}}
\psfrag{b}[c][c][0.8]{\shortstack{ref. ant.\\excited at $f_\text{1}$}}
\psfrag{P}[l][c][0.8]{Port-3}
\psfrag{c}[l][c][0.8]{\shortstack{RX\\Port-2}}
\psfrag{h}[l][c][0.8]{\shortstack{$P_{12}$}}
\psfrag{W}[l][c][0.9]{\shortstack{measurement at\\TX port at $f_\text{0}$}}
\psfrag{M}[c][c][0.8]{$f_\text{0}==1.7$~GHz}
\psfrag{A}[c][c][0.8]{1.88}
\psfrag{B}[c][c][0.8]{$f_\text{1}=1.92$}
\psfrag{C}[c][c][0.8]{1.97}
\psfrag{D}[l][c][0.8]{2~GHz}
\includegraphics[width=0.9\columnwidth]{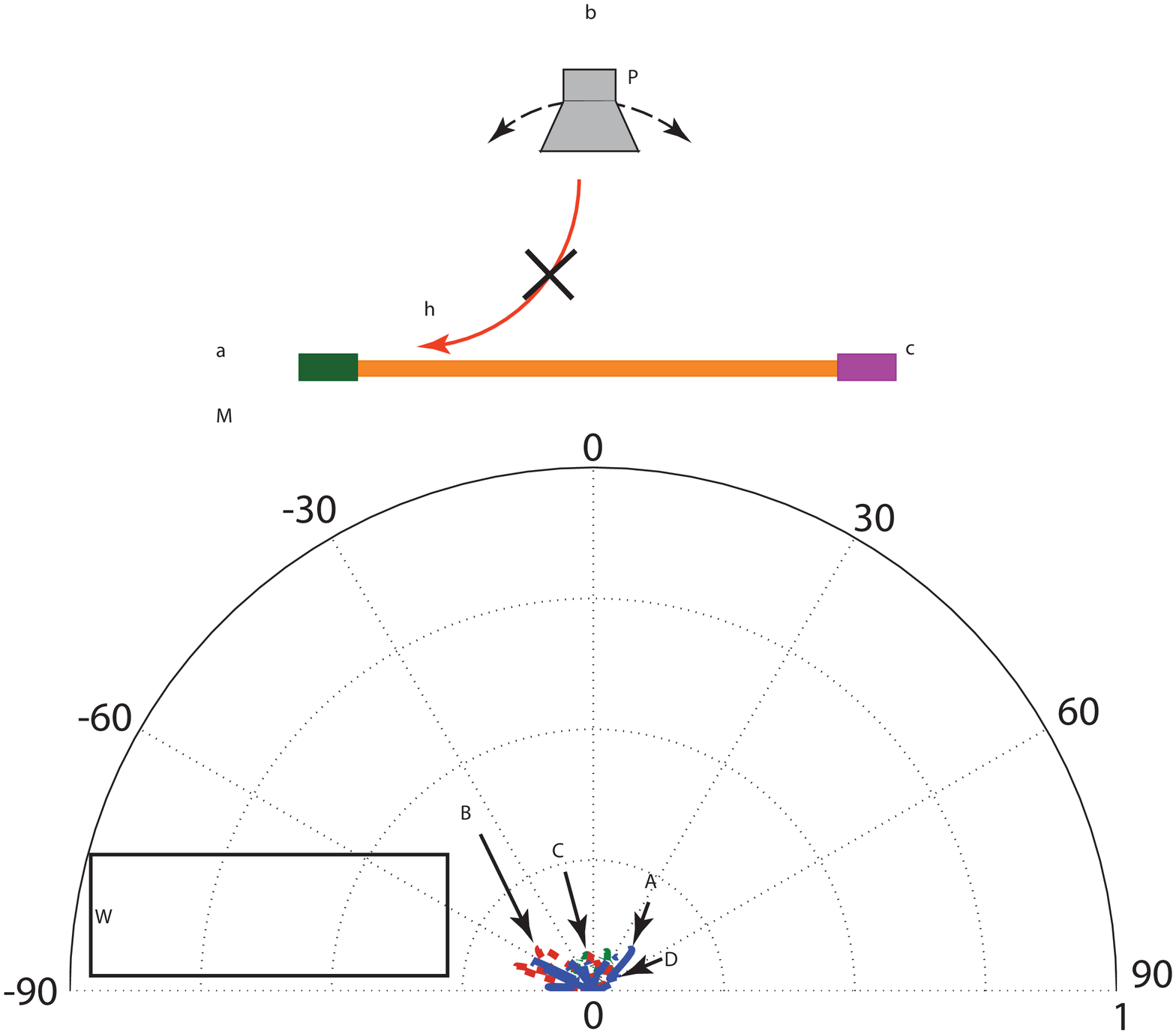}}
\subfigure[]{\label{Fig:res_rec_scanning-c}
\psfrag{A}[c][c][0.8]{$\theta$ (degree)}
\psfrag{B}[c][c][0.8]{\shortstack{Isolation between TX and RX ports\\in receive mode}}
\psfrag{C}[l][c][0.8]{$f_\text{1}=2$~GHz}
\psfrag{D}[l][c][0.8]{$f_\text{1}=1.97$~GHz}
\psfrag{E}[l][c][0.8]{$f_\text{1}=1.92$~GHz}
\psfrag{F}[l][c][0.8]{$f_\text{1}=1.88$~GHz}
\includegraphics[width=0.9\columnwidth]{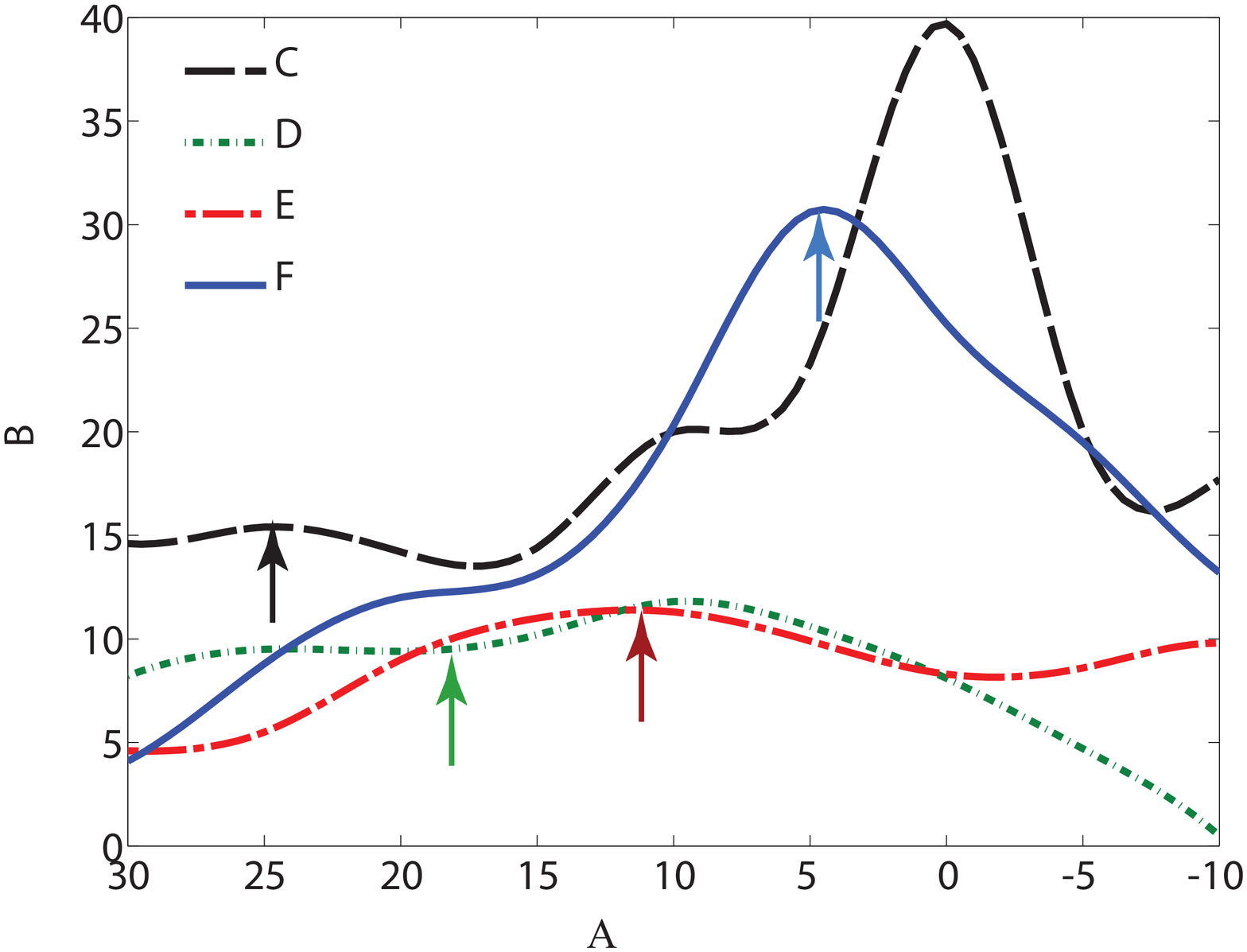}}
\caption{Downlink frequency beam scanning; Normalized received power at the RX and TX ports of the antenna $f_\text{0}=1.7$~GHz for different incoming wave frequency $f_\text{1}=\{1.88, 1.92, 1.97, 2\}$~GHz corresponding to $f_\text{m}=\{0.18, 0.22, 0.27, 0.3\}$~GHz.}
\end{center}
\end{figure}

Figures~\ref{Fig:res_rec_scanning-a} and~\ref{Fig:res_rec_scanning-b} show the experimental normalized received power at the receive and transmit ports for downlink frequency beam scanning. The input frequency at the reference antenna port varies as $f_\text{1}=\{1.88, 1.92, 1.97, 2\}$~GHz, corresponding to $f_\text{m}=\{0.18, 0.22, 0.27, 0.3\}$~GHz, for a fixed received signal $f_\text{0}=f_1-f_\text{m}=1.7$~GHz. Finally, Fig.~\ref{Fig:res_rec_scanning-c} plots the isolation between received powers at the receive and transmit ports.

\section{Conclusion}\label{sec:conc}

We have presented a mixer-duplexer-antenna leaky-wave nonreciprocal system based on periodic space-time modulation using an array of sub-wavelengthly spaced varactors excited by a harmonic wave. The uplink and downlink conversions and duplexing are based on oblique directional space-time transitions from a microstrip leaky mode to itself. One of the interesting features of the system is its capability to perform beam scanning at a fixed signal frequency by varying modulation parameters, in particular the modulation frequency. The theoretical predictions have been verified by experimental demonstration. The proposed system may find applications in various communication, radar and instrument applications.

\appendices

\section{Field and Dispersion Relation Derivation}\label{App_gen}
  \renewcommand{\theequation}{A-\arabic{equation}}
  \setcounter{equation}{0}  

This section derives the electromagnetic field and dispersion relation following a procedure similar to that used (but not detailed) in~\cite{Oliner_PIEEE_1963}.

Due to space-time periodicity, the permittivity can be expanded in the space-time Fourier series,
\begin{equation}
\epsilon (z,t) = \sum\limits_{p =  - \infty }^{+\infty}  {{\epsilon _p}{e^{jp({\omega _{\text{m}}}t - {\beta _{\text{m}}}z)}}},
\label{eqa:perm_fouri}
\end{equation}
while the electric field can be expanded in the space-time Floquet series
\begin{equation}
E(z,t)=\sum_{n =  - \infty}^{+\infty}   E_n e^{j \omega_n t}    e^{-(\alpha_n+j \beta_n) z},
\label{eqa:B-F_sol}
\end{equation}
where $\omega_n= \omega_\text{0} + n\omega _\text{m}$ and $\beta_n(\omega)=\beta_0 (\omega)+n\beta_\text{m}$, with $\beta_0 (\omega)$ and the $E_n$'s being the unknowns to be found. We will first find the coefficients $E_n$, to determine the field in \eqref{eqa:B-F_sol}, and then, based on this result, determine $\beta_0 (\omega)$, which provides the dispersion relation.

We start with the wave equation,
\begin{equation}
\frac{{{\partial ^2}E(z,t)}}{{\partial {z^2}}} - \frac{1}{{{c^2}}}\frac{{{\partial ^2} \left(\epsilon (z,t)E(z,t) \right)}}{{\partial {t^2}}} = 0,
\label{eqa:wave_eq}
\end{equation}
where, using~\eqref{eqa:perm_fouri} and~\eqref{eqa:B-F_sol}, the product under the double time derivative operator reads
\begin{multline}
\epsilon (z,t)E(z,t) \\
 =\sum\limits_{n,p =  - \infty }^{+\infty}   {{\epsilon_p}} {E_n}{e^{j ({\omega_0} + (n + p){\omega_{\text{m}}})t} } e^{-(\alpha_n+j (\beta_0+ (n + p) \beta_\text{m})) z}.
\label{eqa:eps_E_prod}
\end{multline}
For a sinusoidally space-time modulated permittivity, $\epsilon(z,t)= \epsilon_\text{e}(1+\delta_\text{m}\cos(\omega_\text{m}t-\beta_\text{m}z))$ [Eq.~\eqref{eqa:S-T_perm}], Eq.~\eqref{eqa:perm_fouri} reduces to
\begin{equation}
\epsilon (z,t) =  \epsilon_\text{e}\left( \frac{\delta_\text{m}}{2} {e^{ - j({\omega _{\text{m}}}t - {\beta _{\text{m}}}z)}} + 1 + \frac{\delta_\text{m}}{2}{e^{ + j({\omega _{\text{m}}}t - {\beta _{\text{m}}}z)}}\right).
\label{eqa:perm_b}
\end{equation}
Inserting~\eqref{eqa:perm_b} into~\eqref{eqa:eps_E_prod}, and substituting the result and~\eqref{eqa:B-F_sol} into~\eqref{eqa:wave_eq} yields
\begin{multline}
\sum\limits_{n =  - \infty }^{+\infty} \bigg( E_n e^{j \omega_n t}    e^{-(\alpha_n+j \beta_n) z} ( (\alpha_n+j \beta_n)^2 {c^2} + \epsilon_\text{e} \omega_n^2 ) \\
+\frac{\delta_\text{m}}{2} \omega_{n-1}^2 E_n e^{j \omega_{n-1} t} e^{-(\alpha_n+j \beta_{n-1}) z}\\
+\frac{\delta_\text{m}}{2}\omega_{n+1}^2 E_n e^{j \omega_{n+1} t} e^{-(\alpha_n+j \beta_{n+1}) z}\bigg)=0.
\end{multline}
Considering that the summation runs from $-\infty$ to $+\infty$, this expression may be simplified to
\begin{multline}
\sum\limits_{n =  - \infty }^\infty  e^{j (\omega_n t-(\beta_n-j\alpha_n) z)}    \bigg(((\alpha_n+j \beta_n)^2 {c^2} + \epsilon_\text{e} \omega_n^2 )E_n \\
+\frac{\delta_\text{m}}{2} \omega_{n}^2 (E_{n+1} + E_{n-1} ) \bigg)=0.
\end{multline}
Since this relation must hold for all the values of $z$ and $t$, it may finally be expressed as
\begin{equation}
 E_{n - 1} + b_n E_n + E_{n + 1} = 0, \quad -\infty<n<\infty,
\label{eqa:rec_eqs}
\end{equation}
where
\begin{equation}
b_n=\frac{2}{\delta_\text{m}} \left(1- \frac{(\beta_n-j\alpha_n)^2 }{k_{\text{e}n}^2 }  \right).
\label{eqa:bn}
\end{equation}

where $k_{\text{e}n}=\omega_n \sqrt{\epsilon_\text{e}}/c$. To find the $E_n$'s for $n<0$, we employ the following set of recursive equations from~\eqref{eqa:rec_eqs}:
\begin{subequations}\label{eqa:E_neg_sets}
\begin{align}\label{eqa:E_neg_sets-a}
E_{n - 1} + b_n E_n + E_{n + 1} = 0,&&
\end{align}
\begin{align}\label{eqa:E_neg_sets-b}
E_{n - 2} + b_{n - 1} E_{n - 1} + E_{n} = 0.&&
\end{align}
\end{subequations}
Inserting~\eqref{eqa:E_neg_sets-b} into~\eqref{eqa:E_neg_sets-a} gives
\begin{equation}
E_{n - 1} + b_n \left(-E_{n - 2} - b_{n - 1} E_{n - 1} \right)=- E_{n + 1},
\end{equation}
and multiplying both sides by $E_n$ yields
\begin{multline}
E_n \left(E_{n - 1} + b_n \left(-E_{n - 2} - b_{n - 1} E_{n - 1} \right)\right) \\
=- E_{n + 1} \left(-E_{n - 2} - b_{n - 1} E_{n - 1} \right),
\end{multline}
which can be solved for $E_n$ as
\begin{equation}
E_n = E_{n + 1} \frac{E_{n - 2} + b_{n - 1} E_{n - 1}}{ - b_n \left(E_{n - 2} + b_{n - 1} E_{n - 1} \right) + E_{n - 1} },
\end{equation}
and may be rewritten as
\begin{equation}
E_n = E_{n + 1} \frac{1}{ - b_n + \frac{1}{b_{n - 1}+ \frac{E_{n - 2}}{E_{n - 1}}  }}.
\label{eqa:E_neg_En}
\end{equation}
This relation recursively generalizes to the infinite long division expression
\begin{equation}
E_n = E_{n + 1} \frac{1}{ - b_n + \frac{1}{b_{n - 1}+ \frac{1}{ - b_{n - 2} + \frac{1}{b_{n -3}+\ldots    }}    }}.
\label{eqa:E_neg_bn}
\end{equation}
Similarly, to find $E_n$ for $n>0$, we form from~\eqref{eqa:rec_eqs} the following set of recursive equations:
\begin{subequations}
\begin{align}\label{eqa:E_pos_sets-a}
E_{n - 1} + b_n E_n + E_{n + 1} = 0,&&
\end{align}
\begin{align}\label{eqa:E_pos_sets-b}
E_{n } + b_{n + 1} E_{n + 1} + E_{n+2} = 0.&&
\end{align}
\end{subequations}
Following the same procedure as from in~\eqref{eqa:E_neg_sets} to~\eqref{eqa:E_neg_En} for $n<0$, yields
 \begin{equation}
E_n = E_{n - 1} \frac{1}{ - b_n + \frac{1}{b_{n + 1}+ \frac{E_{n + 2}}{E_{n + 1}}    }},
\label{eqa:E_pos}
\end{equation}
which generalizes to
\begin{equation}
E_n = E_{n - 1} \frac{1}{ - b_n + \frac{1}{b_{n + 1}+ \frac{1}{ - b_{n + 2} + \frac{1}{b_{n +3}+\ldots    }}    }}.
\label{eqa:E_pos_bn}
\end{equation}
Equations~\eqref{eqa:E_neg_bn} and~\eqref{eqa:E_pos_bn} recursively provide the amplitude coefficients of the space-time harmonics of the electric field in~\eqref{eqa:B-F_sol} in terms of the excitation electric field $E_0$ and in terms of the $b_n$ coefficients, which themselves depend on $\delta_\text{m}$, $\omega_\text{m}$ and $\omega_\text{0}$.

Let us now derive the dispersion relation, which essentially corresponds to the unknown parameter $\beta_0$ depending on $\omega$ in~\eqref{eqa:B-F_sol}. First, we write~\eqref{eqa:rec_eqs} for $n=0$, i.e.
\begin{equation}
E_{- 1} + b_0 E_0 + E_{+ 1} = 0.
\label{eqa:rec_disp}
\end{equation}
where $E_{- 1}$ and $E_{+ 1}$ can also be expressed in terms of $E_0$  using~\eqref{eqa:E_neg_bn} for $n=-1$ and~\eqref{eqa:E_pos_bn} for $n=+1$, respectively, transforming~\eqref{eqa:rec_disp} into
\begin{equation}
 \frac{E_{0}}{ - b_{- 1} + \frac{1}{b_{-2}+ \frac{1}{ - b_{-3} + \frac{1}{b_{-4}+ \ldots    }}    }} + b_0 E_0 + \frac{ E_{0}}{ - b_1 + \frac{1}{b_{2}+ \frac{1}{ - b_{3} + \frac{1}{b_{4}+ \ldots    }}    }} = 0.
\end{equation}
Dividing this relation by $E_0$ finally yields
\begin{equation}
 \frac{1}{ - b_{- 1} + \frac{1}{b_{-2}+ \frac{1}{ - b_{-3} + \frac{1}{b_{-4}+ \ldots    }}    }} + b_0  + \frac{ 1}{ - b_1 + \frac{1}{b_{2}+ \frac{1}{ - b_{3} + \frac{1}{b_{4}+ \ldots    }}    }} = 0,
\label{eqa:dispers_relat}
\end{equation}
which is a relation based on the $b_n$'s in~\eqref{eqa:bn}. For a given set of modulation parameters $(\delta_\text{m}, \epsilon_\text{e}, \omega_\text{m}, \beta_\text{m}, \alpha_n)$ and variables $\omega_0$ and $\beta_0$, Eq.~\eqref{eqa:dispers_relat} provides the dispersion diagram of the system, $\beta(\omega)=\beta_0(\omega)+n\beta_\text{m}$. The resolution is performed numerically, setting $\omega_0$ and finding the corresponding $\beta_0$ (or vice-versa), and repeating the operation across the frequency range of interest.

\section{Uplink and Downlink Conversions}\label{App_up_down_conv}

For weak modulation, $\delta_\text{m}\ll 1$, the wave amplitudes $E_n$ in ~\eqref{eqa:E_neg_bn} and~\eqref{eqa:E_pos_bn} for $\mid n \mid >1$ are negligible. Moreover, since the leaky-wave antenna is designed to support radiation mostly at $\omega _\text{r}=\omega _\text{s}+\omega _\text{m}$, corresponding to the higher harmonic $E_{+1}$, the amplitude of the lower harmonic $E_{-1}$, corresponding to $\omega _\text{s}-\omega _\text{m}$, can be neglected. Then, the system of equations~\eqref{eqa:rec_eqs} reduces to
 \begin{subequations}\label{eqa:up_E}
\begin{equation}\label{eqa:up_E_a}
 b_0 E_0 + E_{+1} = 0,
\end{equation}
\begin{equation}\label{eqa:up_E_b}
E_{0} + b_{1} E_{+1} = 0,
\end{equation}
\end{subequations}
which has non-trivial solution only if
 \begin{equation}\label{eqa:up_disp_rel_b0b1}
b_0 b_1=1.
\end{equation}
Solving~\eqref{eqa:up_disp_rel_b0b1} with~\eqref{eqa:bn} for $\alpha_n\ll \beta_n$ and $\delta_\text{m}^2 \rightarrow 0$ (weak modulation) yields
\begin{equation}\label{eqa:up_disp_rel}
\beta_\text{0}=\pm \beta_\text{um} \pm \frac{\delta_\text{m}}{4} \sqrt{\beta_\text{um} \beta_\text{um}'},
\end{equation}
where the upper (positive) signs are for forward ($+z$) propagation and the lower (negative) signs are for backward ($-z$) propagation, $\beta_\text{um}'=\beta_\text{um}+\beta_\text{m}$, and $\beta_\text{um}$ is given by~\eqref{eqa:beta_unmod}. Then, using~\eqref{eqa:up_E_b},
\begin{equation}
E_{1} =  \frac{\delta_\text{m} E_0 \beta_\text{um}'}{\delta_\text{m} \sqrt{\beta_\text{um} \beta_\text{um}'}-2\alpha_1^2/\beta_\text{um}'-j \alpha_1\left(\delta_\text{m}  \sqrt{\beta_\text{um}/\beta_\text{um}'}+4 \right)},
\end{equation}
and the uplink power conversion gain reads then
\begin{equation}
\begin{split}
\frac{P_{+1}}{P_0}
&=  \left| \frac{E_{+1}}{E_0} \right|^2\\
= &\left| \frac{\delta_\text{m}^2  \beta_\text{um}'^2 }  {(\delta_\text{m} \sqrt{\beta_\text{um} \beta_\text{um}'}-2\alpha_1^2/\beta_\text{um}')^2+\alpha_1^2 \left(\delta_\text{m}  \sqrt{\beta_\text{um}/\beta_\text{um}'}+4 \right)^2  }\right|.
\end{split}
\end{equation}
Note that in the absence of radiation ($\alpha_1=0$), this expression would reduce to
\begin{equation}
\frac{P_{+1}}{P_0} =  \left|\frac{\beta_\text{um}+\beta_\text{m}}{\beta_\text{um}}\right|,
\label{eqa:trans_conv}
\end{equation}
which is the Manley-Rowe relation~\cite{Manley_Rowe_1956,Manley_1960}, verifying the conservation of energy in the space-time modulated system, and indicating power gain since $\beta_\text{m}>0$.

A similar procedure provides the corresponding results for the downlink using~\eqref{eqa:up_E_a}. The results are
\begin{equation}
E_{0} =  \frac{\delta_\text{m} E_1 \beta_\text{um}}{\delta_\text{m} \sqrt{\beta_\text{um} \beta_\text{um}'}-2\alpha_0^2/\beta_\text{um}-j \alpha_0\left(\delta_\text{m}  \sqrt{\beta_\text{um}'/\beta_\text{um}}+4 \right)},
\end{equation}
and
\begin{equation}
\begin{split}
\frac{P_0}{P_{+1}}
&=  \left| \frac{E_0}{E_{+1}} \right|^2\\
= &\left| \frac{\delta_\text{m}^2  \beta_\text{um}^2 }  {(\delta_\text{m} \sqrt{\beta_\text{um} \beta_\text{um}'}-2\alpha_0^2/\beta_\text{um})^2+\alpha_0^2 \left(\delta_\text{m}  \sqrt{\beta_\text{um}'/\beta_\text{um}}+4 \right)^2  }\right|,
\end{split}
\end{equation}
reducing to the Manley-Rowe relation
\begin{equation}
\frac{P_0}{P_{+1}} =  \left|\frac{\beta_\text{um}}{\beta_\text{um}+\beta_\text{m}}\right|
\label{eqa:rec_conv}
\end{equation}
since $\alpha_0=0$, indicating power loss ($\beta_\text{m}>0$).

\bibliographystyle{IEEEtran}
\bibliography{Travati_ST_LWA_Reference}

\end{document}